\documentstyle[sprocl]{article}
\input{psfig}
\def\Journal#1#2#3#4{{#1} {\bf #2}, #3 (#4)}

\def\NPB{{\em Nucl. Phys.} B}
\def\PLB{{\em Phys. Lett.}  B}
\def\PRL{\em Phys. Rev. Lett.}
\def\PRD{{\em Phys. Rev.} D}

\def\be{\begin{equation}}
\def\ee{\end{equation}}
\def\bea{\begin{eqnarray}}
\def\eea{\end{eqnarray}}

\begin{document}
\title{Like Sign Top Pair Production at LHC\ \footnote{In collaboration with W. S. Hou, C. Y. Ma and C. P. Yuan.} 
}
\author{ Guey-Lin Lin }
\address{Institute of Physics, National Chiao Tung University,
Hsinchu, Taiwan, R.O.C.
}
\maketitle\abstracts{Having a mass comparable to the weak scale, the top quark may have 
a sizable flavor changing couplings to Higgs bosons.
We show that such couplings
can be probed at the LHC through the parton subprocess
$c(\bar c)g \to t (\bar t)A^0$, where the pseudoscalar $A^0$
subsequently decays into $t \bar c$ or ${\bar t} c$,
giving rise to the intriguing final state of like sign top quark pairs.
We also discuss major
backgrounds to the above signal, in particular the QCD-Weak process $q\bar q' \to 
W^+(W^-) t \bar t$. The issue of  background reduction is briefly  discussed.      
}
\section{Introduction}
Despite its excellent agreements with experiments,  the Standard Model  (SM)
offers limited insight into its own structure. 
In particular,  it does not explain but simply parametrizes the hierarchical patterns
seen in both  the fermion masses and the CKM mixing  matrix.
Nor does it reveal any details of the Higgs sector
which is responsible for the electroweak symmetry breaking.
Before one establishes a theory to account for all of  these,  
it is important to gather experimental hints by exploring
the properties of  the top quark and the Higgs boson(s). 
Since $m_t$ is comparable to the weak scale,  
flavor dynamics involving the top quark and 
the  electroweak symmetry breaking mechanism
might be closely related to each other.  
In this talk, we discuss the case where the top quark possesses
large flavor changing couplings to neutral Higgs bosons \cite{Hou,HW},
in a scenario that 
electroweak symmetry breaking is driven by a scalar sector.  
Specifically we will use  the two-Higgs doublet model(2HDM) as an illustration.  
In this model, it is 
customary 
to impose discrete symmetries \cite{GW} to ensure the absence of 
flavor changing neutral Higgs couplings (FCNH) at the tree level. 
However, inspired by the quark mass and mixing hierarchy pattern,
\begin{equation}
   \begin{array}{ccccccc}
    m_1 & \ll & m_2 & \ll & m_3, && \\
    \vert V_{ub}\vert^2 & \ll & \vert V_{cb} \vert^2 &
                      \ll & \vert V_{us}\vert^2 & \ll &1,
   \end{array}
\end{equation}
Cheng and Sher \cite{CS} suggested that low energy FCNC 
can be naturally suppressed without invoking discrete symmetries. 
Hence tree level FCNH couplings are allowed, and particularly those involve top and charm 
quarks are sizable as will be shown later. In the next section, 
we shall discuss the ansatz of Cheng and Sher and its implications on the structure  of  the two-Higgs doublet model.  In section 3, we show that  FCNH couplings may be probed at LHC via
$c(\bar c)g \to t (\bar t)A^0$. Possible backgrounds will  also be identified.
Section 4 is the conclusion.  
\section{The Model} 
The Lagrangian for a  general two-Higgs doublet model (2HDM) can be 
written as 
$L=L_K \ + \ L_Y$,
with
\begin{equation}
L_K=(D_{\mu}\Phi_1)^{\dagger}(D^{\mu}\Phi_1)+(D_{\mu}\Phi_2)^{\dagger}(D^{\mu}\Phi_2)
-V(\Phi_1,\Phi_2),
\end{equation}
and
\begin{eqnarray}
L_Y&=& \lambda^{U(1)}_{ij}\bar{Q}_{iL}\tilde{\Phi}_1U_{jR}+
\lambda^{D(1)}_{ij}\bar{Q}_{iL}\Phi_1D_{jR} \nonumber \\
&+& \lambda^{U(2)}_{ij}\bar{Q}_{iL}\tilde{\Phi}_2U_{jR}+
\lambda^{D(2)}_{ij}\bar{Q}_{iL}\Phi_2D_{jR} \ + \mbox{\ H.c.},
\end{eqnarray}
where $\Phi_{1(2)}$ is the Higgs doublet and $\tilde{\Phi}_{1(2)}$ is its conjugate; 
$Q_{iL}$ denotes the left-handed quark doublet while right-handed quarks are denoted 
by $U_{jR}$ and $D_{jR}$ respectively. $L_Y$ gives rise to quark mass terms 
\begin{eqnarray}
L_M &=& \bar{U}_{iL}(M^{U(1)}_{ij} + M^{U(2)}_{ij} )U_{jR}\nonumber\\
&+& \bar{D}_{iL}(M^{D(1)}_{ij} + M^{D(2)}_{ij} )D_{jR} \ + \mbox{\ H.c.},
\end{eqnarray}
where, $M^{U(k)}_{ij}={\lambda^{U(k)}_{ij} v_k\over \sqrt{2}}$ and
$M^{D(k)}_{ij}={\lambda^{D(k)}_{ij} v_k\over \sqrt{2}}$,
with $v_{1(2)}$ being 
the vacuum expectation values of neutral Higgs. To produce a pattern like  (1), unless fine-tuned cancellations occur,
the off-diagonal elements of  $M^{(1)}$ and $M^{(2)}$, just like those in
their sum $M=M^{(1)}+M^{(2)}$, should trickle off as one moves off-diagonal.\footnote{We refer to both up and down quark cases whenever  the superscripts $U$ and $D$ are not shown.} 
Hence, the FCNH coupling matrices $\xi^{(k)}$,
obtained from $\sqrt{2}M^{(k)}/v_k$ 
by rotating to the mass eigenbasis, cannot be arbitrary.  
Based upon this observation, Cheng and Sher proposed \cite{CS} 
the ansatz      
$\xi^{\left(k\right)}_{ij} \sim {\sqrt{m_i m_j}/v_k}$.
Thus, FCNH couplings involving lower generation fermions 
are naturally suppressed, without pushing
FCNH  Higgs boson masses to way beyond the v.e.v. scale. 
However, since $\sqrt{2} m_t\cong v\equiv \sqrt{v_1^2 + v_2^2}$,  
the flavor  changing coupling  $\xi^{U(k)}_{tc}$ could be quite sizable, 
and could hence lead to interesting consequences
such as $t\to  S^0+c $ \cite{Hou,HW} or $S^0\to t\bar c,\ \bar tc$ \cite{Hou},
where $S^0$ is some neutral Higgs boson. 
Since top decay seems to proceed predominantly via $t\to bW^+$,
we shall concentrate on the case where neutral Higgs bosons
are heavier than the top quark.

Before we proceed to discuss how to probe $\xi^{U(k)}_{tc}$ 
 let us  rotate $\Phi_1$ and $\Phi_2$ \cite{LS} 
such that  $\left\langle \phi_2^0 \right\rangle =0$ and
$\left\langle \phi_1^0 \right\rangle =v/ \sqrt{2}$. This redefinition of fields is 
legitimate because we impose no discrete symmetries on the Lagrangian.
Such a rotation leaves $L_K$ the same while transforms $L_Y$ into:
\begin{eqnarray}
L_Y= &\ - \ & \left( \bar U_L \tilde{M}^U U_R + \bar D_L\tilde{M}^D D_R \right)\,
            \sqrt{2}\, {\mbox{Re}\, \phi_1^0\over v} \nonumber \\
    &\ + \ & \left(\bar U_{L} \xi^{\left(U\right)} U_{R}
               + \bar D_{L} \xi^{\left(D\right)} D_{R}\right)\,
                      \sqrt{2}\, \mbox{Re}\, \phi_2^0 \nonumber \\
      &\ + \ & \left(-\bar U_{L} \xi^{\left(U\right)} U_{R}
               + \bar D_{L} \xi^{\left(D\right)} D_{R}\right)\,
                 i\,\sqrt{2}\, \mbox{Im}\, \phi_2^0 \nonumber \\
   &\ - \ & \bar D_{L} V^\dagger \xi^{\left(U\right)} U_{R}\,  \sqrt{2}\, \phi_2^-
       + \bar U_{L} V\,       \xi^{\left(D\right)} D_{R}\, \sqrt{2}\, \phi_2^+
                \ + \mbox{\ H.c.},
\end{eqnarray}
where $\tilde{M}^{U}$ and $\tilde{M}^{D}$ are diagonal mass matrices.  Note that 
 FCNH couplings $\xi^{U(D)}$ are now only associated with the second doublet, and they
are just linear combinations of  those in the old basis. Therefore, for both up and down-quark
case
 \begin{equation}
\xi_{ij} =f_{ij} {\sqrt{m_i m_j}/v},
\end{equation}
where $f_{ij}$'s are constants of order unity. 
In this new basis, the pseudoscalar $A^0 \equiv \sqrt{2}\, \mbox{Im}\, \phi_2^0$
and charged scalar $H^\pm \equiv \phi_2^\pm$ are physical Higgs bosons.
The CP even neutral scalars 
$\sqrt{2}\, \mbox{Re}\,\phi_1^0$ and $\sqrt{2}\, \mbox{Re}\,\phi_2^0$
mix through the Higgs potential into the physical states $H^0$ and $h^0$.
In the limit that the mixing angle $\sin\alpha \to 0$,
$H^0 \leadsto \sqrt{2}\, \mbox{Re}\, \phi_1^0$
becomes the ``standard" Higgs boson with
diagonal couplings,
while
$h^0 \leadsto \sqrt{2}\, \mbox{Re}\, \phi_2^0$
has Yukawa couplings as in   (5),
but decouples from vector
or $H^+H^-$ boson pairs, just like $A^0$.
\section{Signatures of  FCNH Couplings at LHC} 
Recently there are already discussions on how to probe 
$\xi_{tc}$ coupling 
at $\ell^+\ell^-$ colliders 
\cite{ARS,HANDL}. 
In particular,  it was suggested that \cite{HANDL}
this coupling can be probed at  a 500GeV Next Linear Collider through the production $e^+e^- \to Z^*\to h^0 A^0$  with $h^0$ and $A^0$ each decaying into $t\bar c(\bar {t} c)$ to form an intriguing like sign top final state.  In the mass range  
 \begin{equation}
200\ \mbox{GeV} < m_{h^0,\; A^0} < 2m_t \simeq 350\ \mbox{GeV},
\end{equation}
and the limit  $\sin\alpha\to 0$, this final state is most  favorable since both
$h^0$ and $A^0$ predominantly decay into $t\bar{ c }(\bar{t} c)$.  With a 50fb$^{-1}$ integrated luminosity,  one expects a score of  like sign dilepton events
in NLC annually. For a generic $\sin\alpha$, $h^0$ mainly decays into massive gauge
bosons, and like sign dilepton events still occur following the decays of  $W^+W^-t\bar{c}(\bar{t}c)$ final state. Once again, one expects a few like sign dilepton events each year.  Note that $\xi_{tc}$ is not  directly probed by this process, since it enters only in decays of $h^0$ and $A^0$. To look for 
\begin{figure} 
\psfig{figure=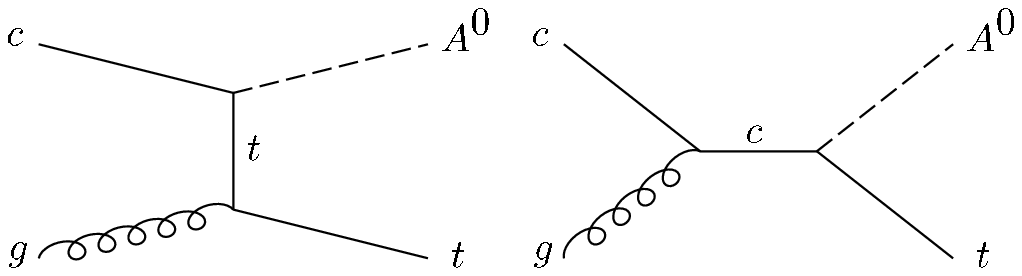,height=1.27in} \caption{Subprocess $cg\to tA^0$.}
\end{figure}
a direct probe 
on $\xi_{tc}$ with a larger
event rate, we 
turn to hadronic colliders.  In subsequent discussions, we shall again focus on the mass range 
given by eq. (7). 
 
The advantage of hadronic colliders lies in their capabilities of involving 
the strong interaction in the {\it production process}, 
which can be used to {\it directly probe  $\xi_{tc}$}.  
Surveying  $q\bar{q},\ qg$ and $gg$ processes,
to have $\xi_{tc}$ appearing in one of the interaction vertices,
one in general encounters $2\to 3$ scattering, such as
$q\bar{q}\to g^*\to t\bar{c}(\bar{t} c)A^0$.  
However, the cross sections turn out to be very small,
and it would be advantageous if $2\to 2$ scattering is possible.
We find that the $c(\bar {c})g\to t(\bar{t})A^0$ process
is rather promising in this regard as a direct probe to  $\xi_{tc}$. 
Although its cross section at Tevatron remains small, 
the situation changes drastically at LHC.

At Tevatron energies, to produce an $A^0$ of 250 GeV 
in association with a top quark, the colliding partons must carry
large momentum fractions, hence both charm and gluon 
distribution functions are very suppressed, 
resulting in a very small $tA^0$ production cross section. 
From  Fig. 1 and using CTEQ3L \cite{CTEQ3} parton distribution functions, 
the cross section at the Tevatron is only about $10^{-2} f^2 $ fb for $m_{A^0}=250$ GeV,
where $f=f_{tc}$ is the constant appearing in   (6). 
Though the cross section is very small, it is proportional to $f^2$
hence a direct probe to FCNH coupling $\xi_{tc}$.  
At the LHC with $\sqrt{s}=14$ TeV, 
the colliding parton momentum fractions could be much smaller 
so that both charm and gluon distribution functions contribute significantly. 
Repeating the calculation for LHC, with $m_{A^0}$= 250 GeV
we obtain a cross section of $37f^2$ fb which is 
3000 times larger than that at the Tevatron,
and grows as $f^2$.
We show in Fig. 2 the dependence of  $\sigma (pp\to t(\bar{t})A^0+X)$
on $m_{A^0}$ with $f$ taken to be unity.  

We note that \cite{HANDL}
$A^0$ decays predominantly into $t\bar{c}$ or $\bar{t}c$ 
in the mass range given by   (7).
For example, for $m_{A^0}=250$ GeV, 
$90\%$ (fraction increases with $m_{A^0}$) of  $A^0$ 
decays into the above final states \cite{HANDL}, half of which 
pair up with the 
\begin{figure} 
\psfig{figure=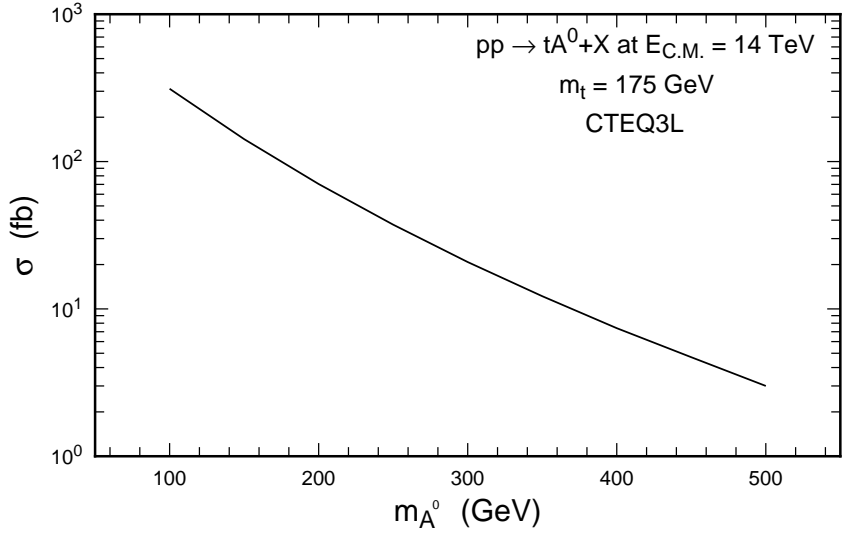,height=2.8in} \caption{Cross section for $pp\to tA^0 + X$ at LHC via subprocess of Fig. 1.}
\end{figure}
associated top to make a like sign top pair event.
The signature of such events are like sign dileptons,
accompanied by two $b$-jets, large missing energy,
plus one additional jet,
\begin{equation}
cg\to tA^0\to \ell_1^+\ell_2^+\nu\nu + bb + \bar{c},
\end{equation}
and similarly for
$\bar{c}g\to \bar{t}A^0\to \ell_1^-\ell_2^- \bar{\nu}\bar{\nu} + \bar{b}\bar{b} + c$.
With an integrated luminosity of 100 fb$^{-1}$ 
and 50$\%$ double $b$-tagging efficiency,
we expect for both $\ell^+\ell^+$ and $\ell^-\ell^-$ modes
\begin{equation}
 37f^2\times {90\%\over 2}\times {4\over 81} \times 50\% \times 100=40f^2
\end{equation} 
events per year for $m_{A^0}=250$ GeV.
The event rate for other values of $m_{A^0}$ can be read off from Fig. 2,
together with Fig. 1 of  Ref. 7, 
where the $m_{A^0}$ dependence of  BR($A^0\to t\bar{c}+\bar{t} c$) is plotted. 
Over the mass range of   (7), the event rate does not change significantly
since $m_{A^0}$ affects the production cross section 
and BR($A^0\to t\bar{c}+\bar{t} c$) in compensating ways.

One might think that the same final state may also be reached by 
single top production followed by $A^0$ bremsstrahlung, as shown in Fig. 3. 
This is analogous to the production of a Higgs boson 
associated with a single top \cite{WG}. 
In the current  context, we have
\begin{equation}
qb\to q' t A^0.
\end{equation} 
\begin{figure} 
\psfig{figure=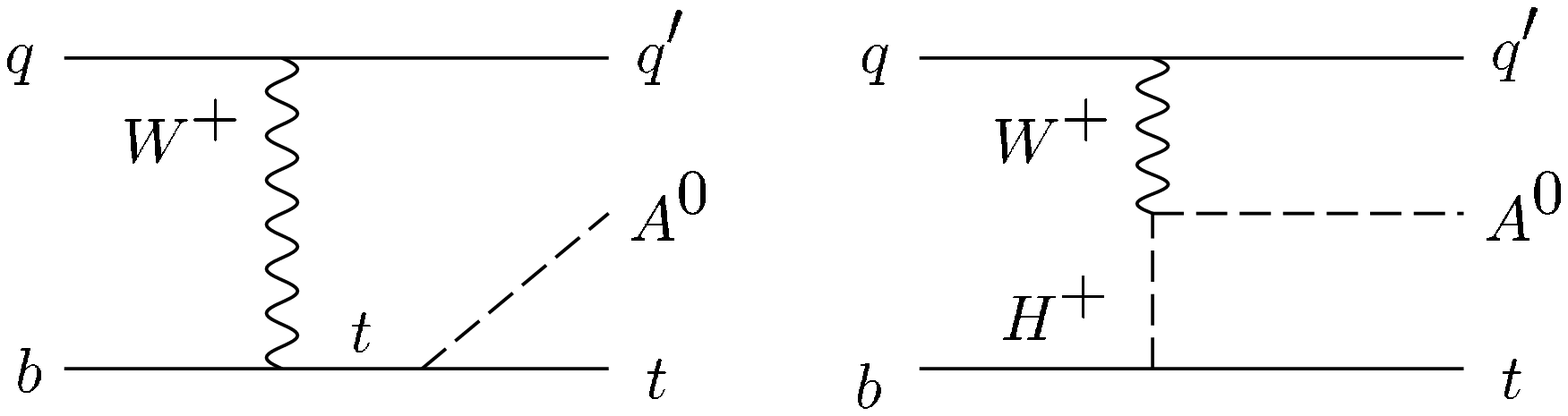,height=1.25in} \caption{Subprocess $qb\to q^\prime t A^0$.
}
\end{figure}
Since the simpler parent process, 
the so-called single-top production $qb\to q't$,
has a cross section around 100 pb \cite{CY},
the process (10) would appear to have a  large cross section.  
Adding an $A^0$ to the final state tends to 
reduce the cross section by 3 orders of magnitude, 
but a cross section for $qb\to q' t A^0$ around 100 fb is still quite large.

To ascertain this, we divide the total cross section into three parts,
$\sigma =\sigma_t + \sigma_{H^+} +\sigma_{tH^+}$,
where the first two terms are from each diagram alone,
and the third is their interference.
With CTEQ3L \cite{CTEQ3} parton distribution functions,
we find $\sigma_t$, $\sigma_{H^+}$,  and $\sigma_{tH^+}$ 
to be 21.7, 24.4 and  $-$43.6 fb, respectively, 
for $m_{H^+} = m_{A^0} = $ 250 GeV. 
The interference term almost cancels the diagonal terms completely
and renders a total cross section of only 2.5 fb, 
which is much smaller than that of $cg \to tA^0$! 
The result is found to be not very sensitive to $m_{H^+}$, and was double checked with helicity methods \cite{CY}.
That such a 
cancellation must occur is due to the requirement of unitarity. Upon a closer inspection on both
diagrams,  one can easily identify their correlations.
First, both  $A^0t t$ and  $H^+bt $
couplings are proportional to $m_t$,  and both $Wbt$ and $WHA^0$ couplings are  proportional to the weak coupling constant $g$.  Second, denominators of  top and charged Higgs 
propagators have opposite signs, which, along with correlations in coupling constants,    
lead to the cancellation of two diagrams.  

We have now singled out $cg\to tA^0$ as the most promising  
mode to probe the FCNH coupling $\xi_{tc}$.  
This is a {\it direct probe of $\xi_{tc}$},
since $f$ can be determined from the cross section of $cg\to tA^0$. 
What remains to be checked are the backgrounds. 
We focus on like sign $W$ pair production,
which would also give rise to  like sign dilepton events.

Vector boson pair production has been studied extensively \cite{BB}
for the purpose of  probing the electroweak symmetry breaking mechanisms. 
By requiring two $b$-jets and like sign dileptons in the final state, 
one can discard all of these,
\begin{figure} 
\psfig{figure=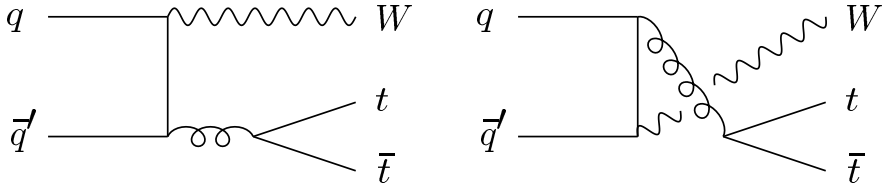,height=0.95in} \caption{Standard model $q\bar q^\prime \to Wt\bar t$ subprocess..
}
\end{figure}
{\it except} 
$q\bar{q'} \to W^+(W^-)t\bar{t}$ \cite{KUN} (see Fig. 4).  
The production and decay chain
\begin{equation}
u\bar{d}\to W^+ t\bar{t}\to W^+W^+W^-b\bar{b}
\to \ell_1^+\ell_2^+\nu\nu + b\bar{b} + j_1j_2,
\end{equation}
leads to like sign dileptons
as well as a pair of $b$- and $\bar b$-jets
(and likewise for 
$d\bar{u}\to W^- t\bar{t}\to W^-W^-W^+b\bar{b}\to
\ell_1^-\ell_2^-\bar{\nu}\bar{\nu} + b\bar{b} + j_1j_2$).
Unlike the signal process of (8), there are {\it two jets} $j_1$ and $j_2$,
which should have pair mass $m_{jj}$ around $M_W$.
Convoluting with parton distribution functions, 
we find $\sigma ( pp \to W^+ t\bar{t} +X)=210$ fb 
while $\sigma (pp \to W^- t\bar{t} +X)=100$ fb, 
which agrees with the results of Barger {\em et al.} in 
Ref. 12.

Assuming an integrated luminosity100 fb$^{-1}$ at the LHC, 
the annual event number for process (11) is
\begin{equation}
210\times {2\over 3}\times {4\over 81}\times 50\% \times 100=350,
\end{equation}
and half this rate for $\ell^-\ell^- + X$ events.
The factor  of ${2/3}$ is the $W\to jj$ branching ratio.  
The background of (12) appears to dominate over the signal of (9) 
both in $\ell^+\ell^+$ and $\ell^-\ell^-$ modes, 
though it is less severe in the latter case.  
Adding to the problem, we find that the $W$ boson associated with the $t\bar t$ pair  
also turns out to be produced in the central region,
hence a Monte Carlo study is needed 
to separate signal from background.
While details of such a study will be presented elsewhere, 
let us provide a qualitative argument on this matter. 
The simplest  way is clearly jet counting. 
Two $b$-jets are already tagged, 
but it may be too costly to determine $b$ vs. $\bar b$.
The signal has one additional jet  
while the background has two, with $m_{jj} \simeq M_W$. 
If the two background jets are both in the central region 
($\vert\eta\vert < 3$, where $\eta$ is pseudorapidity) 
and can be distinguished, 
the event can be excluded by na\"\i ve jet counting. 
If the two jets merge into one large jet $J$, 
the event can still be effectively removed
by cutting on large $m_J$ around $M_W$. 
Only if either $j_1$ or $j_2$ falls outside of the detection region
or coalesce accidentally with one of the $b$-jets
will the event become an irreducible background. 
This kinematics is however unlikely because the $t\bar t$ syetem,
which gives rise to $W\to j_1 j_2$, is centrally produced
as discussed before.
A conservative estimate is that by  jet counting  alone, 
one should be able to reduce the background by at least 50$\%$ \cite{GPY}.

With simple jet counting, one has less than 90   
background $\ell^-\ell^-$ events, 
with an excess of  $\sim 40f^2$ coming from signal events. 
For $f\sim \sqrt{2}$, signal and background event rates
would be comparable.
In other words, if  the FCNH coupling 
$\xi_{tc} = f {\sqrt{m_c m_t}/v}$ indeed exists, 
considerable excess shall be observed in $\ell^-\ell^-$ events.
For a slightly larger $f$, say $f\sim 2$, the signal is also 
comparable to the background in the  $\ell^+\ell^+$ mode. 
Defining $N(\ell^\pm\ell^\pm)$ as the number of  $\ell^\pm\ell^\pm$ events, 
the signal can then manifest itself in the asymmetry parameter
\begin{equation}
 A={N(\ell^+\ell^+)-N( \ell^-\ell^-)\over N(\ell^+\ell^+)+N( \ell^-\ell^-) }.
\end{equation} 
The background alone gives $A={1\over 3}$, 
while the signal events lower $A$ to ${1\over 7}$ for $f\sim 2$.  

Before closing this section, we wish to point out that  $cg\to tA^0$ can be viewed as 
a model independent probe to FCNH couplings. 
To produce like sign top events, only a sizable $A^0tc$ coupling $\xi_{tc}$ 
and a fairly large branching ratio for  $A^0 \to t\bar{c}+\bar{t}c$ are essential. 
If  the first condition is satisfied, it is very probable that the second one holds as well.  
One simply needs to argue that  the branching ratio for two boson decays,
$A^0\to VV$, is suppressed.  
In fact, assuming CP conservation, the $A^0VV$  couplings exist only in dimension 5 operators or beyond. Hence, in a renormalizable theory, $A^0VV$ coupling is indeed suppressed since it can only be generated by loop corrections.   
\section{Conclusion}
In summary, we have shown that FCNH couplings can be directly probed at the LHC 
via like sign top quark pair production through the $cg\to tA^0\to tt\bar q$ process. 
Possible backgrounds are identified and calculated.
To better distinguish the signal from backgrounds,  a detailed Monte Carlo study is called for.  
Furthermore it is also essential  to calculate 
the background $q\bar{q'}\to W^\pm t\bar{t}$ more accurately, 
in particular to the NLO accuracy where dependence on 
the renormalization (factorization) scale could be significantly reduced.   
\section{Acknowledgments}
I  like to give thanks to  Chao-Qiang Geng, Wei-Ming Zhang and others for their efforts in organizing this workshop, and to Darwin Chang, Jyh-Liong Lim and Gong-Ping Yeh  for useful discussions.
This work is supported in part by 
National Science Council of R.O.C. under the grant number 
NSC 86-2112-M-009-012.

\end{document}